\pdfoutput=1

\documentclass[11pt,a4paper]{article}
\usepackage{jheppub}

\usepackage[english]{babel}
\usepackage{graphicx}
 
\usepackage{amsmath,amsfonts,amssymb}
\usepackage{dsfont}      
\usepackage{enumerate}
\usepackage{array}

\newcommand\Lagr{\mathcal{L}}
\newcommand\identity{\mathds{1}}

\DeclareMathOperator{\tr}{Tr}

\title{Classical skyrmions in SU(N)/SO(N) cosets}

\author{Marc Gillioz}

\affiliation{Institute for Theoretical Physics, University of Zurich, \\
	Winterthurerstrasse 190, CH-8057 Z\"urich, Switzerland}

\emailAdd{gillioz@physik.uzh.ch}

\proceeding{\vspace{-1.5cm}\flushright{\small ZU-TH 05/11 \\ LPN11-15 \\}}

\abstract{We construct the skyrmion solutions appearing in the coset spaces \linebreak $SU(N)/SO(N)$ for $N > 2$ and compute their classical mass. For $N=3$, the third homotopy group $\pi_3(SU(3)/SO(3)) = \mathbb Z_4$ implies the existence of two distinct solutions: the skyrmion of winding number two has spherical symmetry and is found to be the lightest non-trivial field configuration; the skyrmion and antiskyrmion of winding number plus and minus one are slightly heavier and of toroidal shape. For $N \geq 4$, there is only one skyrmion since the third homotopy group is $\mathbb Z_2$. It is found to have spherical symmetry and is significantly lighter than the $N=3$ solutions.}

\notoc

\begin{document}

\maketitle

\setcounter{page}{0}

\vspace{5cm}

\flushbottom

\section{Introduction}

Skyrmions~\cite{Skyrme:1961vq} are topological solitons appearing in field theories where the set of mappings from three-dimensional space $\mathbb R^3$ to a target space $\mathcal M$ is split into non-equivalent homotopy classes, characterised by the third homotopy group $\pi_3(\mathcal M)$. The lowest energy configuration in each equivalence class cannot be transformed continuously to the vacuum and is therefore stable over time. If those field configurations are degenerate in energy, they represent different vacua of the theory. If not, only the lightest one is a true vacuum state, the others being called skyrmions and representing upon quantisation distinct particle states.

In the low-energy chiral theory of QCD, $\sigma$-models describing mesons are known to contain skyrmions, which represent the baryons of the theory~\cite{Witten:1983tx,Adkins:1983ya}. In this case the global symmetry group consists of flavour $SU(2)$, or can be extended to flavour $SU(3)$. The non-trivial homotopy group $\pi_3(SU(N)) = \mathbb Z$ allows the existence of an infinite tower of skyrmions characterised by an integer charge, the baryon number. Baryons as skyrmions of a scalar theory of mesons have been extensively studied in the literature (see~\cite{Weigel:2008zz} for a review). Recent results include stabilisation of the skyrmion from a spatial extra-dimension~\cite{Pomarol:2007kr,Domenech:2010aq}.

Many modern theories trying to solve the hierarchy problem raised by the Higgs mechanism of electroweak symmetry breaking enlarge the particle content and the global symmetry group of the Standard Model. Some of them predict new scalar fields --- both elementary or appearing as fermion condensates --- at low or medium energy scales, and share therefore similar features with the Skyrme model. This is the case in particular in models where the Higgs boson is a pseudo-Nambu-Goldstone boson of some strongly interacting theory, as in the composite Higgs models~\cite{Kaplan:1983fs,Kaplan:1983sm}, or more recently in little Higgs models~\cite{ArkaniHamed:2001nc,ArkaniHamed:2002qx,ArkaniHamed:2002qy,Low:2002ws,Schmaltz:2004de,Freitas:2009jq,Schmaltz:2010ac}, as well as in other strongly interacting composite models arising from a holographic description~\cite{Contino:2003ve,Agashe:2004rs,Giudice:2007fh}. All these models describe the Nambu-Goldstone sector where the Higgs lives in terms of a $\sigma$-model, with various symmetry groups depending on the exact realisation of the model. Some of these symmetry group have a trivial topology and can not lead to the existence of skyrmions, but other do.

In this paper we will focus on the case of a $SU(N)/SO(N)$ $\sigma$-model, which has a non-trivial topology as characterised by its third homotopy group~\cite{Bryan:1993hz}
\begin{equation}
	\pi_3(SU(N)/SO(N)) = \mathbb Z_p,
	\hspace{1.5cm}
	p = \left\{\begin{array}{lll}
		4 & ~ \textrm{for} ~ & N = 3, \\
		2 & ~ \textrm{for} ~ & N \geq 4.
	\end{array}\right.
	\label{eq:homotopy}
\end{equation}
Such a symmetry is realised with $N=5$ in the littlest Higgs model~\cite{ArkaniHamed:2002qy}, for which skyrmions are present~\cite{Hill:2007zv} and can even act as dark matter~\cite{Murayama:2009nj,Joseph:2009bq,Gillioz:2010mr}.\footnote{The non-trivial topology of the $SU(5)/SO(5)$ coset also induces the presence of other topological defects in the littlest Higgs model, such as cosmic strings and $\mathbb Z_2$ monopoles~\cite{Trodden:2004ea}.} In general, the $SU(N)/SO(N)$ symmetry breaking pattern can be obtained from a strongly interacting theory with Weyl fermions in the adjoint representation of the gauge group~\cite{Auzzi:2006ns,Bolognesi:2007ut,Auzzi:2008hu,Bolognesi:2009vm}. The finiteness of the third homotopy groups~(\ref{eq:homotopy}) implies that $SU(N)/SO(N)$ skyrmions have quite different physical properties from their $SU(N)$ cousins. Spherically symmetric skyrmion solutions of winding number $B=2$ for $N=3$ and of winding number $B=1$ for $N \geq 4$ have been constructed before in ref.~\cite{Balachandran:1982ty,Balachandran:1983dj} and~\cite{Gillioz:2010mr} respectively. We review these two cases and complete the analysis by constructing a solution of winding number $B = \pm 1$ for $N=3$, whose symmetry is only axial.

This work is organised as follows: in section two the model is introduced and general considerations about $SU(N)/SO(N)$ skyrmions are presented. In the third section we present a general method to construct solution of unit winding number and apply it to the cases $N=3$ and $N \geq 4$. The classical masses of the corresponding skyrmions are computed. The fourth section contains the construction of the $N=3$ skyrmions of winding number two. Finally, in the fifth section, the results are summarised and further issues are shortly discussed.


\section{SU(N)/SO(N) skyrmions}

We consider a $\sigma$-model defined by the Lagrangian density
\begin{eqnarray}
	\Lagr & = & \Lagr_2 + \Lagr_4, \nonumber \\
	& = & \frac{f^2}{4} \tr \partial_\mu \Sigma \partial^\mu \Sigma^\dag
		+ \frac{1}{32 e^2} \tr \left[ \Sigma^\dag \partial_\mu \Sigma, \Sigma^\dag \partial_\nu \Sigma \right]
			\left[ \Sigma^\dag \partial^\mu \Sigma, \Sigma^\dag \partial^\nu \Sigma \right],
	\label{eq:lagrangian}
\end{eqnarray}
where $\Sigma(x)$ is a $SU(N)$ \emph{symmetric} matrix. The model has a global $SU(N)$ symmetry under which $\Sigma(x)$ transforms in the two-indices symmetric representation as
\begin{equation}
	\Sigma \mapsto U \Sigma U^T,
	\hspace{1cm}
	U \in SU(N).
	\label{eq:symmetry}
\end{equation}
The vacuum state is obtained by taking a constant value for the field $\Sigma(x)$. Here we choose it to be the identity matrix
\begin{equation}
	\langle \Sigma \rangle = \identity_N.
	\label{eq:vev}
\end{equation}
This vacuum expectation value spontaneously breaks the global $SU(N)$ symmetry down to $SO(N)$, since only the $SO(N)$ subgroup of $SU(N)$ leaves the vacuum unchanged under the transformation rule~(\ref{eq:symmetry}). As a consequence the Nambu-Goldstone field $\Sigma(x)$ takes its values in the coset $SU(N)/SO(N)$.

The Lagrangian~(\ref{eq:lagrangian}) is identical to the one of the Skyrme model~\cite{Skyrme:1961vq} describing the low-energy chiral limit of QCD, in which $f$ is identified with the pion decay constant and $e$ is a parameter depending on the high-energy behaviour of the theory, found empirically to be $e \cong 5$~\cite{Adkins:1983ya}. The notable difference is that in the original Skyrme model the field transforms in the adjoint representation of $SU(N_f)_L \times SU(N_f)_R$, where $N_f$ is the number of light flavours, and a global diagonal $SU(N_f)_V$ symmetry is preserved after spontaneous symmetry breaking.

The presence of skyrmions in the model~(\ref{eq:lagrangian}) is due to the fact that the third homotopy group $\pi_3$ of the target space is non-trivial. However, this condition is not sufficient to ensure the existence of skyrmions. If the four-derivative term $\Lagr_4$ were omitted, the mass and size of the skyrmions would shrink to zero, according to Derrick's theorem~\cite{Derrick:1964ww}: under a rescaling of the spacetime coordinates $x \to \lambda \, x$, the energy associated with the kinetic term scales as $E_2 \to \lambda \, E_2$ and can thus be made arbitrarily small. Only the interplay of $\Lagr_2$ and $\Lagr_4$, whose energy scales as $E_4 \to (\lambda)^{-1} E_4$, ensures a stable size for the skyrmion. In the presence of both terms, the energy is bounded by a minimum $E_\textrm{min} = 2 \sqrt{E_2 E_4}$, reached when $E_2$ and $E_4$ contribute equally.

In this paper, for simplicity, we will always consider static solutions --- moving \linebreak skyrmions can be obtained by applying a Lorentz boost --- and work in dimensionless units $\tilde x = (f e) \, x$. With this choice, the energy density corresponding to the Lagrangian~(\ref{eq:lagrangian}) is
\begin{equation}
	E[\Sigma] = \frac{f}{e} \int d^3\tilde x \, \tr \left[ \frac{1}{4} \partial_i \Sigma \partial_i \Sigma^\dag
		+ \frac{1}{16} \left( \partial_i \Sigma \partial_i \Sigma^\dag \partial_j \Sigma \partial_j \Sigma^\dag
		- \partial_i \Sigma \partial_j \Sigma^\dag \partial_i \Sigma \partial_j \Sigma^\dag \right) \right].
	\label{eq:energy}
\end{equation}
The integrand is then independent of the parameters $f$ and $e$ and can be directly computed from the shape of the skyrmion solution.

The elements $\Sigma(x)$ of the coset $SU(N)/SO(N)$ can be parametrised by the Cartan embedding~\cite{Bryan:1993hz,Gillioz:2010mr,Auzzi:2006ns}, starting from a matrix $\Phi(x)$ in $SU(N)$, as
\begin{equation}
	\begin{array}{ccc}
		SU(N) & \to & SU(N)/SO(N), \\
		\Phi(x) & \mapsto & \Sigma(x) = \Phi(x) \Phi(x)^T.
	\end{array}
	\label{eq:cartanembedding}
\end{equation}
The matrix $\Phi(x)$ is only defined up to right-multiplication with an $SO(N)$ matrix. The advantage of this embedding is that the winding number of a field configuration $\Phi(x)$ in $SU(N)$ can be simply expressed in terms of an integral over space, as\footnote{The notation $B$ comes from the original Skyrme model of mesons, in which the winding number is identified with the conserved baryon number of the theory.}
\begin{equation}
	B(\Phi) = \frac{1}{24 \pi^2} \epsilon_{ijk} \int d^3 x~ \tr (\Phi^\dag \partial_i \Phi)
		(\Phi^\dag \partial_j \Phi) (\Phi^\dag \partial_k \Phi)
		~ \in \mathbb Z.
	\label{eq:windingnumber}
\end{equation}
This integral is a topological invariant, and is additive with respect to multiplication of two fields $\Phi_1(x)$ and $\Phi_2(x)$:
\begin{equation}
	B(\Phi_1 \Phi_2) = B(\Phi_1) + B(\Phi_2) + \frac{1}{8 \pi^2} \int d^3 x ~
		\partial_i \Omega_i,
\end{equation}
where $\Omega_i = \epsilon_{ijk} \tr ( \Phi_1^\dag \partial_j \Phi_1 ) ( \Phi_2 \partial_k \Phi_2^\dag )$. 
The last term is a surface term and therefore vanishes provided that the fields are in the vacuum at the spatial boundaries, $\Phi_{1,2}(x \to \infty) = \identity$.

The winding number of $\Sigma(x)$ can then be straightforwardly identified with the winding number of the field $\Phi(x)$ used to construct it. However, this mapping is not unique: the multiplication of $\Phi(x)$ from the right with an $SO(N)$ matrix $R(x)$ leaves $\Sigma(x)$ unchanged, but not the winding number~(\ref{eq:windingnumber}), which is raised or lowered as $B(\Phi R) = B(\Phi) + B(R)$. The crucial point here is that $B(R)$ cannot take an arbitrary integer value, since the winding number integral~(\ref{eq:windingnumber}) evaluated on an element of $SO(N)$ gives an integer multiple of 4 for $N=3$ and of 2 for $N \geq 4$~\cite{Bryan:1993hz}. Therefore the winding number of an element in the coset $SU(N)/SO(N)$ is only defined modulo a factor of 4 or 2, for $N=3$ and $N \geq 4$ respectively. Moreover, two field configuration $\Sigma_1(x)$ and $\Sigma_2(x)$ built respectively from $SU(N)$ fields $\Phi_1(x)$ and $\Phi_2(x)$ differing by one unit of winding number cannot be continuously transformed into another and thus belong to different homotopy classes. This is an immediate consequence of the statement made above that $\pi_3(SU(3)/SO(3)) = \mathbb Z_4$ and $\pi_3(SU(N)/SO(N)) = \mathbb Z_2$ for $N \geq 4$. The winding number for the field $\Sigma(x) = \Phi(x)\Phi(x)^T$ can thus be expressed unambiguously as
\begin{equation}
	B(\Sigma) = B(\Phi) ~ \textrm{mod} ~ p,
\end{equation}
where $p$ is defined as in eq.~(\ref{eq:homotopy}).

With the help of the Cartan embedding~(\ref{eq:cartanembedding}), the challenge of constructing a field configuration in each of the homotopy classes of the coset space reduces to finding an appropriate configuration $\Phi(x)$ in $SU(N)$ with the desired winding number. A naive choice consists in taking the lightest $SU(N)$ skyrmion solution of a given winding number and building the corresponding $SU(N)/SO(N)$-valued field directly out of it. For $B = 1$ the lightest solution is known to be the hedgehog configuration~(\ref{eq:hedgehog}) with spherical symmetry and mass $M_0 = 72.9~\frac{f}{e}$ with our conventions~\cite{Adkins:1983ya}. The Cartan embedding of this solution yields a skyrmion of mass $4 M_0 = 291.7~\frac{f}{e}$ whose symmetry is however only axial. This is a hint that the hedgehog ansatz might not yield the lightest field configuration, as we shall see. For $B = 2$, the $SU(N)$ skyrmion has a toroidal shape, and its mass of $139.6~\frac{f}{e}$ is slightly lighter than twice the mass of a single skyrmion~\cite{Braaten:1988cc}. The naive Cartan embedding of this $B=2$ configuration gives a mass of $599.0~\frac{f}{e}$ for the corresponding $SU(N)/SO(N)$ field. We can therefore infer the following upper bounds on the skyrmion masses:
\begin{equation}
	M_{B = \pm 1} \leq 291.7 ~ \frac{f}{e},
	\hspace{1cm}
	M_{B = 2} \leq 599.0 ~ \frac{f}{e}.
	\label{eq:upperbound}
\end{equation}
We are actually going to show in the next sections that the true skyrmion solutions have masses much below these naive bounds.

Notice finally that in the original $SU(N)$ Skyrme model a six derivative term is often present as an alternative to the four-derivative term $\Lagr_4$ or in addition to it. It contains only two time derivatives and preserves all the symmetries of the Skyrme Lagrangian~\cite{Adkins:1983nw,Jackson:1985yz}:
\begin{equation}
	\Lagr_6 = c_6 \tr B_\mu B^\mu,
	\label{eq:L6}
\end{equation}
where $B^\mu$ is the topological current defined as 
\begin{equation}
	B^\mu = \frac{1}{24 \pi^2} \epsilon^{\mu\nu\rho\sigma} \int d^3 x~ \tr (\Phi^\dag \partial_\nu \Phi)
		(\Phi^\dag \partial_\rho \Phi) (\Phi^\dag \partial_\sigma \Phi).
	\label{eq:topologicalcurrent}
\end{equation}
However, the equivalent of this current with the field $\Sigma(x)$ replacing $\Phi(x)$ vanishes identically due to the structure of the coset, so that the Lagrangian term~(\ref{eq:L6}) is automatically absent of the theory. For the same reason, the usual Bogomolny bound~\cite{Bogomolny:1975de} for the mass of a $SU(N)$ skyrmion of winding number $B$, $M_B \geq 6 \pi^2 |B| \frac{f}{e}$, does not apply since the component $B_0$ of the topological current~(\ref{eq:topologicalcurrent}) is zero. A crucial consequence is that the mass of the skyrmion is not necessarily increasing with increasing winding number. Indeed, we shall see in the following that for $N=3$ the $B=2$ skyrmion is lighter than the $B = \pm 1$ ones.


\section[The solutions of winding number B = 1]{The solutions of winding number \boldmath $B = 1$}

The skyrmion configuration of unit winding number in $SU(N)$ has been known for a long time. It is obtained by using a so-called hedgehog ansatz living in a $SU(2)$ subgroup of $SU(N)$. Due to the symmetries of the original Skyrme Lagrangian, the embedding of the $SU(2)$ subgroup can be chosen without loss of generality to be in the upper-left $2 \times 2$ block of the $SU(N)$ matrix as
\begin{equation}
	\Phi(x) = \left( \begin{array}{cc}
		\Phi_0(x) & \\ & \identity_{N-2}
	\end{array} \right),
	\label{eq:ansatz:B1}
\end{equation}
where $\Phi_0(x)$ is the previously mentionned hedgehog ansatz described by
\begin{equation}
	\Phi_0(x) = \exp\left[i \, F(r) \, \hat x_i \, \sigma_i \right].
	\label{eq:hedgehog}
\end{equation}
Here $F(r)$ is a function of the radial coordinate, $\hat x_i = x_i / r$ are angular coordinates, and the $\sigma_i$ are the usual Pauli matrices. The boundary condition for $F$ at spatial infinity is fixed to $F(\infty) = 0$ in order to recover the vacuum $\Phi(x \to \infty) = \identity$, while for definiteness at the origin $F(0)$ has to be an integer multiple of $\pi$. Chosing $F(0) = \pi$ ensures the correct unit winding number.

The hedgehog ansatz is built so that a spatial $SO(3)$ rotation of the coordinates $x_i \to R_{ij} x_j$ is equivalent to a $SU(2)$ transformation $\Phi_0 \to U_0 \Phi_0 U_0^\dag$, with the equivalence between the two given by $R_{ij} = \frac{1}{2} \tr(\sigma_i U_0 \sigma_j U_0^\dag)$. Since the original Skyrme Lagrangian is symmetric under diagonal $SU(N)$ transformations --- hence under any $SU(2)$ subgroup --- it is also invariant under spatial rotations. For this reason, the skyrmion built using this hedgehog ansatz is said to be \emph{spherically symmetric}. In general, soliton configurations with the highest symmetry tend to have the lowest energy: this is true not only for skyrmions, but also for monopoles and dyons. We want therefore to follow the same general principles to construct the $SU(N)/SO(N)$ skyrmion.

The first important point about the $SU(N)$ skyrmion construction is that it only makes use of a $SU(2)$ subgroup of the target space. This is the minimal choice, as proven in a theorem due to Bott~\cite{Bott:1956tf,Bryan:1993hz}: the winding number integral~(\ref{eq:windingnumber}) is actually only counting the number of times the field is winding around any $SU(2)$ subgroup of $SU(N)$,\footnote{This is a direct consequence of the fact that the physical space $\mathbb R^3$ is isomorphic to $S^3 \sim SU(2)$ upon identification of the spatial infinity to a single point on the sphere.} which makes it sufficient to embed all the non-trivial components of the field $\Phi(x)$ into a $SU(2)$ subgroup, while the other can be taken in the vacuum. We postulate here that this rule can also be applied to the construction of the $SU(N)/SO(N)$ skyrmion, so that in general the field $\Phi(x)$ can be written as $\Phi(x) = \exp[i \, f_i(x) T_i]$, where the $T_i$ are generators of a $SU(2)$ subgroup of $SU(N)$.

However, while in the original Skyrme model the relevant subgroup could be trivially chosen to live in the upper-left $2 \times 2$ block of the matrix $\Phi$, this is not the case in the coset, where not every choice of $SU(2)$ subgroup is equivalent: among the diagonal $SU(N)$ transformations $\Phi \to U \Phi U^\dag$, only the ones satisfying $U^\dag \langle \Sigma \rangle = \langle \Sigma \rangle U^T$ are symmetries of the model. The form of the ansatz~(\ref{eq:ansatz:B1}) is nevertheless very useful to ensure a correct winding number by construction, so instead of considering a different ansatz for $\Phi$ we keep this form and work in a different basis where the vacuum expectation value is not necessarily diagonal. The most general form of the Cartan embedding~(\ref{eq:cartanembedding}) is $\Sigma(x) = \Phi(x) \Sigma_0 \Phi(x)^T$, where $\Sigma_0 = \langle\Sigma\rangle$ denotes the vacuum state.

The second crucial point about the construction of the $SU(N)$ skyrmion is to impose spherical symmetry. We have seen that this is possible if the  transformation $\Phi_0 \to U_0 \Phi_0 U_0^\dag$ is a symmetry of the Lagrangian. This $SU(2)$ transformation of $\Phi_0$ can be promoted to a $SU(N)$ transformation by considering $\Phi \to U \Phi U^\dag$, with
\begin{equation}
	U = \left( \begin{array}{cc}
		U_0 & \\ & V
	\end{array} \right),
	\hspace{1cm}
	U_0 \in SU(2),
	\quad
	V \in SU(N-2).
	\label{eq:U:definition}
\end{equation}
As discussed above, this transformation is a symmetry of the Lagrangian only if it satisfies
\begin{equation}
	U^\dag \Sigma_0 = \Sigma_0 U^T.
	\label{eq:U:commutation}
\end{equation}
Since $U_0^\dag$ and $U_0^T$ live respectively on the left- and right-hand side of eq.~(\ref{eq:U:commutation}) and cannot be related by multiplication with a matrix $\Sigma_0$ for arbitrary values of $U_0$, the equation can only be fulfilled if the $(N-2)$-dimensional matrix $V$ is proportional to $U_0^*$. From this point, we have to consider two separated cases:
\begin{enumerate}
\item
For $N = 3$, the one-dimensional matrix $V$ is fixed to $V = 1$, hence eq.~(\ref{eq:U:commutation}) cannot be satisfied and the $SU(2)$ transformation rule of $\Phi_0$ cannot be promoted to a symmetry of the Lagrangian, independently of the basis. Thus we conclude that there exist no spherically symmetric ansatz of unit winding number in the coset $SU(3)/SO(3)$.
\item
On the contrary, for $N \geq 4$, choosing
\begin{equation}
	\Sigma_0 = \left( \begin{array}{ccc}
		& \identity_2 & \\ \identity_2 && \\ && \identity_{N-4}
	\end{array} \right),
	\hspace{1cm}
	U = \left( \begin{array}{ccc}
		U_0 && \\ & U_0^* & \\ && \identity_{N-4}
	\end{array} \right),
	\label{eq:vacuum:N4}
\end{equation}
the transformation $\Phi \to U \Phi U^\dag$ acts on $\Phi_0$ as a diagonal $SU(2)$ transformation $\Phi_0 \to U_0 \Phi_0 U_0^\dag$ and simultaneously preserves the vacuum, so that the Lagrangian can be made spherically symmetric. This is the most general choice of basis, up to global $SU(N)$ transformations, and we expect therefore to build the lightest skyrmion configuration with the help of this ansatz.
\end{enumerate}

\subsection[The $N \geq 4$ skyrmion]{The \boldmath $N \geq 4$ skyrmion}

\begin{figure}
	\centering
	\includegraphics[width=0.47\linewidth]{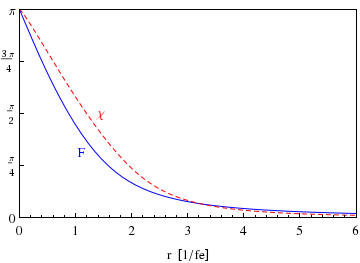}
	\includegraphics[width=0.51\linewidth]{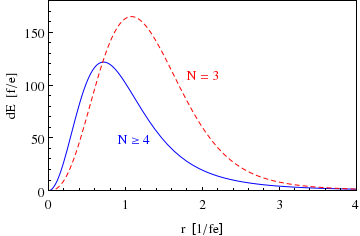}
	\caption{Left: The profile functions $F$ and $\chi$ for the $N=4$, $B=1$ and $N=3$, $B=2$ skyrmions respectively. Right: the corresponding radial energy distributions $dE = -4\pi \Lagr$ (right) as functions of the radius $r$ in dimensionful units of $1/f e$.}
	\label{fig:F-chi-E}
\end{figure}

With the choice of basis~(\ref{eq:vacuum:N4}), we have just proven that the skyrmion can be made spherically symmetric. Indeed, an important property of the Skyrme Lagrangian~(\ref{eq:lagrangian}) is that it can be expressed in terms of the currents $\Sigma^\dag \partial_\mu \Sigma$, which take in this case a block diagonal form,
\begin{equation}
	\Sigma = \left( \begin{array}{ccc}
		& \Phi_0 & \\ \Phi_0^T && \\ && \identity_{N-4}
	\end{array} \right)
	\quad \Rightarrow \quad
	\Sigma^\dag \partial_\mu \Sigma = 
	\left( \begin{array}{ccc}
		-\left( \Phi_0 \partial_\mu \Phi_0^\dag \right)^T && \\ & \Phi_0^\dag \partial_\mu \Phi_0 & \\ && ~0_{N-4}
	\end{array} \right),
\end{equation}
so that each of the two $2 \times 2$ blocks contributes equally to the energy density~(\ref{eq:energy}), which becomes twice the functional expressed in terms of the $SU(2)$-valued field $\Phi_0(x)$,
\begin{equation}
	E[\Sigma] = 2 \, E[\Phi_0].
\end{equation}
Hence the mass of the $SU(N)/SO(N)$ skyrmion with $N \geq 4$ is exactly twice the mass of the original $SU(2)$ skyrmion. The solution minimising the energy $E[\Phi_0]$ is obtained using the hedgehog ansatz~(\ref{eq:hedgehog}), for which the winding number integral~(\ref{eq:windingnumber}) is unity:
\begin{equation}
	B(\Phi) = -\frac{2}{\pi} \int\limits_0^\infty dr \, F' \, \sin^2 F
		= \left.\frac{\sin F \cos F - F}{\pi}\right|_{F = \pi}^{F = 0} = 1.
\end{equation}
The energy density --- explicitly independent of the angular coordinates due to the spherical symmetry --- reads
\begin{equation}
	E[F] = 4\pi \, \frac{f}{e} \int\limits_0^\infty dr \left[ \left(r^2 + 2 \sin^2 F \right) F'^2
		+ \left( 2 r^2 + \sin^2 F \right) \frac{\sin^2 F}{r^2} \right],
	\label{eq:energy:1}
\end{equation}
and is minimised when $F$ is a solution of the Euler-Lagrange equation
\begin{equation}
	\left( r^2 + 2 \sin^2 F \right) F'' + 2 r F'
		+ \sin 2F \left( F'^2 - 1 - \frac{\sin^2 F}{r^2} \right) = 0.
\end{equation}
The numerical solution of this equation and the corresponding energy density are displayed in Figure~\ref{fig:F-chi-E}. The skyrmion mass computed as the total energy of this field configuration is then
\begin{equation}
	M_{N = 4,B=1} = 145.8 ~ \frac{f}{e}.
	\label{eq:mass:N4B1}
\end{equation}
This mass is exactly twice the mass of the $SU(N)$ skyrmion, and only half the value of the upper bound~(\ref{eq:upperbound}) found from the naive Cartan embedding of the original solution. We do not have a proof that it is indeed the lightest topologically non-trivial field configuration, but it is the only simple embedding preserving the spherical symmetry of the Lagrangian, and it is much lighter than other constructions based on different choices of ansatz, as for example the $N=3$ case discussed below. Note that this solution is the one used in \cite{Gillioz:2010mr} with $N=5$, although the choice of basis $\langle\Sigma\rangle = \identity_5$ makes the ansatz there look more complicated than the simple form~(\ref{eq:ansatz:B1}).

\subsection[The $N=3$ skyrmion]{The \boldmath $N=3$ skyrmion}

The solution of unit winding number in $SU(3)/SO(3)$ was not computed so far in the literature, and requires a more involved ansatz. The reason for this is that the spherical symmetry cannot be preserved, as argued before. 

\begin{figure}
	\centering
	\includegraphics[width=0.48\linewidth]{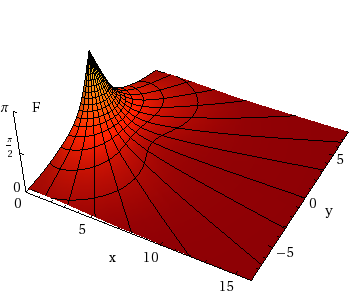}
	\includegraphics[width=0.48\linewidth]{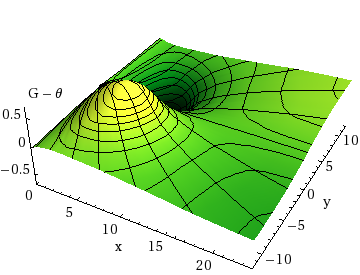}
	\caption{The numerical solution in the case $N=3$, $B=1$ for $F$ (left) and for the difference $G - \theta$ (right) showed in the plane $(x,y) = (\tilde r \sin\theta, \tilde r \cos\theta)$.}
	\label{fig:F2-G2}
\end{figure}
\begin{figure}
	\parbox{0.48\linewidth}{\includegraphics[width=\linewidth]{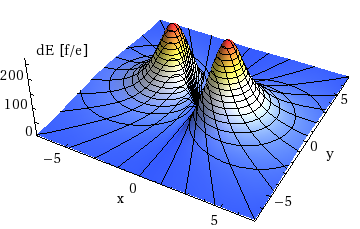}}
	\parbox{0.48\linewidth}{
		\centering \includegraphics[width=\linewidth]{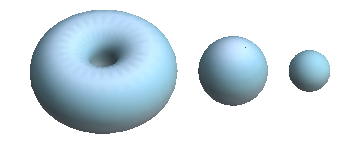} \\
		\begin{tabular}{>{\hspace{1cm}}c>{\hspace{0.7cm}}c>{\hspace{0.3cm}}c}
			$N=3$ & $N=3$ & $N \geq 4$ \\
			$B= \pm 1$ & $B=2$ & $B=1$
		\end{tabular}}
	\caption{Left: The energy density of the $N=3$, $B=1$ solution in a plane containing the axis of symmetry of the skyrmion (located along $x = 0$). Right: isosurfaces of energy density $dE = 100 \, \frac{f}{e}$ showing the relative size of the three solutions.}
	\label{fig:E2}
\end{figure}

In the original basis $\langle\Sigma\rangle = \identity_3$, the ansatz~(\ref{eq:ansatz:B1}) nevertheless allows to preserve an axial symmetry in the Lagrangian, and we expect therefore that it yields the lowest-energy skyrmion solution. The most general form of $\Phi_0$ preserving this axial symmetry can then be written in analogy to the construction of the deuteron solution~\cite{Braaten:1988cc} as
\begin{equation}
	\Phi_0(x) = \exp\left[ i \, F(r, \theta) \, n_i(r,\theta,\varphi) \, \sigma_i \right],
	\label{eq:axialsymmetry}
\end{equation}
where $n$ is a vector of unit length, given in spherical coordinates by
\begin{equation}
	(n_1, n_2, n_3) = \left( \sin G(r,\theta) \sin \varphi, \, \cos G(r,\theta), \, \sin G(r,\theta) \cos \varphi \right).
\end{equation}
Instead of a single profile function $F$ of the radial variable $r$ as in the hedgehog ansatz~(\ref{eq:hedgehog}), we are now left with two functions $F$ and $G$ of two variables $r$ and $\theta$. The boundary conditions are fixed to $F(0, \theta) = \pi$, $F(\infty, \theta) = 0$, $G(r, 0) = 0$ and $G(r, \pi) = \pi$, so that the field $\Phi(x)$ is well defined everywhere. With this ansatz, the winding number integral~(\ref{eq:windingnumber}) becomes
\begin{equation}
	B(\Phi) = \frac{1}{\pi} \int\limits_0^\infty dr \int\limits_0^\pi d\theta ~ \sin^2 F \sin G
		\left( \partial_r F \partial_\theta G - \partial_\theta F \partial_r G \right)
\end{equation}
and gives the expected value of one upon integration by parts. Similarly, a solution of winding number $B=-1$ (or equivalently $B=3$) is obtained by inverting the boundary condition for $F$ at the origin to $F(0,\theta) = -\pi$; the energy of this solution is identical to the $B=1$ case, and we will therefore not discuss it further. Plugging the ansatz~(\ref{eq:axialsymmetry}) into the energy functional (\ref{eq:energy}), one obtains
\begin{eqnarray}
	E[F,G] & = & 4 \pi \int\limits_0^\infty dr \int\limits_0^\pi d\theta \sin\theta
		\left\{ \sin^2 G \left( r^2 + \frac{\sin^2 2F \sin^2 G}{\sin^2 \theta} \right)
		\left[ \left( \partial_r F \right)^2 + \frac{\left( \partial_\theta F \right)^2}{r^2} \right]  \right. \nonumber \\
		&& \hspace{1.5cm} + \sin^2 F \left( r^2(1 - \cos^2 F \sin^2 G) + \frac{\sin^2 F \sin^2 2G}{\sin^2 \theta} \right)
		\left[ \left( \partial_r G \right)^2 + \frac{\left( \partial_\theta G \right)^2}{r^2} \right] \nonumber \\
		&& \hspace{1.5cm} + \frac{1}{2} \sin 2F \sin 2G \left( r^2 + 4 \frac{\sin^2 F \sin^2 G}{\sin^2 \theta} \right)
			\left[ \partial_r F \partial_r G + \frac{\partial_\theta F \partial_\theta G}{r^2}  \right] \nonumber \\
		&& \hspace{1.5cm} + 4 \sin^4 F \sin^2 G ~ \left[ \partial_r F \partial_\theta G - \partial_\theta F \partial_r G \right]^2 \nonumber \\
		&& \hspace{1.5cm} + \left. \frac{\sin^2 F \sin^2 G (1 - \sin^2 F \sin^2 G)}{\sin^2 \theta} \right\}.
	 \label{eq:energy:3}
\end{eqnarray}
The corresponding Euler-Lagrange equations for $F$ and $G$ are too long to be displayed here, but can be solved numerically using relaxation methods. The solution of these equations are shown on Figure~\ref{fig:F2-G2}, and the corresponding energy density on Figure~\ref{fig:E2}. As one can see, unlike the spherical energy distribution of the previous two sections, the energy density is in this case located along a torus.\footnote{Note that our solution resembles the numerical skyrmion solution of the $O(3)$ $\sigma$-model of ref.~\cite{Faddeev:1996zj}.}
The mass of the skyrmion, obtained by integrating this energy density, is found to be
\begin{equation}
	M_{N = 3,B = \pm1} = 273.5 ~ \frac{f}{e}.
	\label{eq:mass:3}
\end{equation}


\section[The solution of winding number $B=2$ for $N=3$]{The solution of winding number \boldmath $B=2$ for \boldmath $N=3$}

The general rule stating that solutions with the highest symmetries tend to have the lowest mass also apply to the $N=3$ case, where the spherically symmetric solution of winding number two found by Balachandran et al.~\cite{Balachandran:1982ty,Balachandran:1983dj} is the lightest known skyrmion configuration. It is constructed from the $SO(3)$ generators $T_i \in \{ \lambda_2, \lambda_5, \lambda_7 \}$, where the $\lambda_i$ denote Gell-Mann matrices. The ansatz is built as follows:
\begin{equation}
	\Phi(x) = \exp\left[ i \, \alpha(r) \, \hat x_i \, T_i \right]
		\exp\left[ i \, \chi(r) \left( \frac{2}{3} \identity - (\hat x_i T_i)^2 \right) \right].
	\label{eq:ansatz:2}
\end{equation}
The two exponentials commute; the first of them belongs to $SO(3)$ and is therefore irrelevant in the coset space, as can be seen from the Cartan embedding~(\ref{eq:cartanembedding}), yielding
\begin{equation}
	\Sigma(x) = \exp\left[ 2 i \, \chi(r) \left( \frac{2}{3} \, \identity - (\hat x_i \, T_i)^2 \right) \right].
\end{equation}
However, it is the first exponential which fixes the winding number~(\ref{eq:windingnumber}):
\begin{equation}
	B(\Phi) = \frac{2}{\pi} \int\limits_0^\infty dr \left[ \alpha' \left( \cos\alpha \cos\chi - 1 \right) 
		- \chi' \, \sin\alpha \sin\chi \right]
		= \left.\frac{2}{\pi} \left( \sin\alpha \cos\chi - \alpha \right)\right|_{r = 0}^{r \to \infty},
\end{equation}
since the expected value $B = 2$ of the winding number is achieved by choosing the boundary conditions $\alpha(\infty) = 0$ and $\alpha(0) = \pi$. Nevertheless, with this choice the first exponential in~(\ref{eq:ansatz:2}) is ill-defined around the origin, and the introduction of the second exponential with boundary conditions $\chi(\infty) = 0$ and $\chi(0) = \pi$ is required to cancel the angular dependence at the boundary point $r = 0$. Plugging this ansatz into the energy density~(\ref{eq:energy}), one obtains the following functional, also independent of the angular variables,
\begin{equation}
	E[\chi] = 8\pi \, \frac{f}{e} \int\limits_0^\infty dr \left[ \left( \frac{1}{3} r^2 + 2 \sin^2 \chi \right) \chi'^2
		+ \left( 2 r^2 + \sin^2 \chi \right) \frac{\sin^2 \chi}{r^2} \right],
	\label{eq:energy:2}
\end{equation}
which is minimised when $\chi$ satisfies the Euler-Lagrange equation
\begin{equation}
	\left( \frac{1}{3} r^2 + 2 \sin^2 \chi \right) \chi'' + \frac{2}{3} r \chi'
		+ \sin 2\chi \left( \chi'^2 - 1 - \frac{\sin^2 \chi}{r^2} \right) = 0.
\end{equation}
The numerical solution for $\chi(r)$ is shown on Figure~\ref{fig:F-chi-E}, together with the energy distribution along the radial direction. The mass of the $B = 2$ skyrmion is hence
\begin{equation}
	M_{N = 3,B = 2} = 228.7 ~ \frac{f}{e}.
	\label{eq:mass:2}
\end{equation}


\section{Summary and conclusions}

The masses of the three different $SU(N)/SO(N)$ skyrmion solutions are summarised in Table~\ref{tab:masses}. For $N=3$, the skyrmions of winding number plus and minus one are heavier than the one of winding number two, so that the mass hierarchy is inverted compared to $SU(N)$ models. Although this result can be surprising, it could be expected from the symmetry of the solutions. Notice that the two masses are nevertheless close to each other --- only about 20\% difference --- whereas in the $SU(N)$ case the Bogomolny bound~\cite{Bogomolny:1975de} implies that the mass of the second heaviest skyrmion --- which has a toroidal shape as in the $SU(3)/SO(3)$ case --- is nearly twice the mass of the lightest, spherically symmetric skyrmion. The $N \geq 4$ solution with spherical symmetry is significantly lighter than both $N=3$ skyrmions. It is actually a rather surprising fact that for any values of $N > 2$, there always exists a spherically symmetric solution, and that this solution is the lightest even when its winding number is not unity.
\begin{table}
	\centering
	\begin{tabular}{|>{\hspace{0.5cm}}c<{\hspace{0.5cm}}
			|>{\hspace{0.5cm}}c<{\hspace{0.5cm}}
			|>{\hspace{0.5cm}}c<{\hspace{0.5cm}}|}
		\cline{2-3}
		\multicolumn{1}{c|}{} & \boldmath $B = \pm 1$ & \boldmath $B = 2$
		\tabularnewline \hline
		\boldmath $N=3$ & $273.5~\frac{f}{e}$ & $228.7~\frac{f}{e}$
		\tabularnewline \hline
		\boldmath $N \geq 4$ & $145.8~\frac{f}{e}$ & \multicolumn{1}{c}{}
		\tabularnewline \cline{1-2} 
	\end{tabular}
	\caption{Classical masses of the $SU(N)/SO(N)$ skyrmions.}
	\label{tab:masses}
\end{table}

The finiteness of the third homotopy groups for $N=3$ and $N \geq 4$ ensure special annihilation properties for the skyrmions: in the latter case, the skyrmion is its own antiskyrmion, so that any two skyrmions can annihilate into a final state with trivial topology. For $N=3$, the skyrmion of winding number two has the same properties as the $N \geq 4$ one; on the other hand, two toroidal skyrmions can annihilate into a state containing a spherical symmetric skyrmion or not, depending if they have the same or opposite winding number. Notice also that although multiple skyrmion solutions are not favoured by energy considerations, no important attractive force between them is expected at large distances~\cite{Gillioz:2010mr}, so that the skyrmions appearing in theories described at intermediate energies by a $SU(N)/SO(N)$ $\sigma$-model can be long-lived and hence provide viable dark matter candidates~\cite{Murayama:2009nj,Joseph:2009bq,Gillioz:2010mr}. The $N=3$ case is especially interesting since the two different kinds of skyrmion are both stable and can therefore account for a fraction of the currently observed relic density each.

Note also that the masses computed in this work are only obtained at the classical level. The quantisation of the model typically makes the physical masses increase depending on the spin of the skyrmion~\cite{Adkins:1983ya}. The stability of the skyrmion in the quantised version of the theory is also questionable, but recent results proved that it is ensured in strongly coupled models where the symmetry breaking pattern $SU(N)/SO(N)$ arises from a condensate of fermions in the adjoint representation of a $SU(N)$ gauge group~\cite{Auzzi:2006ns,Bolognesi:2007ut,Auzzi:2008hu,Bolognesi:2009vm}. Notice moreover that the inclusion of higher-dimensional terms in the Lagrangian~(\ref{eq:lagrangian}) would also modify the skyrmion masses, but not their symmetry properties nor the way they are embedded in the coset. Finally, realistic models may require to gauge totally or partially the global $SU(N)/SO(N)$ symmetry, which can affect the skyrmion stability and mass~\cite{Gillioz:2010mr}. In these cases, a detailed study of the model is required.


\subsection*{Acknowledgements}

I am thankful to Pedro Schwaller and Andreas von Manteuffel for many useful discussions and comments on the draft.
This work was supported by the Schweizer Nationalfonds and by the European Commission through the ``LHCPhenoNet'' Initial Training Network PITN-GA-2010-264564.


\bibliography{Bibliography}

\providecommand{\href}[2]{#2}\begingroup\raggedright\begin{thebibliography}{10}

\bibitem{Skyrme:1961vq}
T.~H.~R. Skyrme, {\it {A Nonlinear field theory}},  {\em Proc. Roy. Soc. Lond.}
  {\bf A260} (1961) 127--138.

\bibitem{Witten:1983tx}
E.~Witten, {\it {Current Algebra, Baryons, and Quark Confinement}},  {\em Nucl.
  Phys.} {\bf B223} (1983) 433--444.

\bibitem{Adkins:1983ya}
G.~S. Adkins, C.~R. Nappi, and E.~Witten, {\it {Static Properties of Nucleons
  in the Skyrme Model}},  {\em Nucl. Phys.} {\bf B228} (1983) 552.

\bibitem{Weigel:2008zz}
H.~Weigel, {\it {Chiral Soliton Models for Baryons}},  {\em Lect. Notes Phys.}
  {\bf 743} (2008) 1--274.

\bibitem{Pomarol:2007kr}
A.~Pomarol and A.~Wulzer, {\it {Stable skyrmions from extra dimensions}},  {\em
  JHEP} {\bf 03} (2008) 051--051,
  [\href{http://xxx.lanl.gov/abs/0712.3276}{{\tt arXiv:0712.3276}}].

\bibitem{Domenech:2010aq}
O.~Domenech, G.~Panico, and A.~Wulzer, {\it {Massive Pions, Anomalies and
  Baryons in Holographic QCD}},  {\em Nucl. Phys.} {\bf A853} (2011) 97--123,
  [\href{http://xxx.lanl.gov/abs/1009.0711}{{\tt arXiv:1009.0711}}].

\bibitem{Kaplan:1983fs}
D.~B. Kaplan and H.~Georgi, {\it {SU(2) x U(1) Breaking by Vacuum
  Misalignment}},  {\em Phys. Lett.} {\bf B136} (1984) 183.

\bibitem{Kaplan:1983sm}
D.~B. Kaplan, H.~Georgi, and S.~Dimopoulos, {\it {Composite Higgs Scalars}},
  {\em Phys. Lett.} {\bf B136} (1984) 187.

\bibitem{ArkaniHamed:2001nc}
N.~Arkani-Hamed, A.~G. Cohen, and H.~Georgi, {\it {Electroweak symmetry
  breaking from dimensional deconstruction}},  {\em Phys. Lett.} {\bf B513}
  (2001) 232--240, [\href{http://xxx.lanl.gov/abs/hep-ph/0105239}{{\tt
  hep-ph/0105239}}].

\bibitem{ArkaniHamed:2002qx}
N.~Arkani-Hamed {\em et.~al.}, {\it {The Minimal Moose for a Little Higgs}},
  {\em JHEP} {\bf 08} (2002) 021,
  [\href{http://xxx.lanl.gov/abs/hep-ph/0206020}{{\tt hep-ph/0206020}}].

\bibitem{ArkaniHamed:2002qy}
N.~Arkani-Hamed, A.~G. Cohen, E.~Katz, and A.~E. Nelson, {\it {The littlest
  Higgs}},  {\em JHEP} {\bf 07} (2002) 034,
  [\href{http://xxx.lanl.gov/abs/hep-ph/0206021}{{\tt hep-ph/0206021}}].

\bibitem{Low:2002ws}
I.~Low, W.~Skiba, and D.~Tucker-Smith, {\it {Little Higgses from an
  antisymmetric condensate}},  {\em Phys. Rev.} {\bf D66} (2002) 072001,
  [\href{http://xxx.lanl.gov/abs/hep-ph/0207243}{{\tt hep-ph/0207243}}].

\bibitem{Schmaltz:2004de}
M.~Schmaltz, {\it {The simplest little Higgs}},  {\em JHEP} {\bf 08} (2004)
  056, [\href{http://xxx.lanl.gov/abs/hep-ph/0407143}{{\tt hep-ph/0407143}}].

\bibitem{Freitas:2009jq}
A.~Freitas, P.~Schwaller, and D.~Wyler, {\it {A Little Higgs Model with Exact
  Dark Matter Parity}},  {\em JHEP} {\bf 12} (2009) 027,
  [\href{http://xxx.lanl.gov/abs/0906.1816}{{\tt arXiv:0906.1816}}].

\bibitem{Schmaltz:2010ac}
M.~Schmaltz, D.~Stolarski, and J.~Thaler, {\it {The Bestest Little Higgs}},
  {\em JHEP} {\bf 09} (2010) 018,
  [\href{http://xxx.lanl.gov/abs/1006.1356}{{\tt arXiv:1006.1356}}].

\bibitem{Contino:2003ve}
R.~Contino, Y.~Nomura, and A.~Pomarol, {\it {Higgs as a holographic
  pseudo-Goldstone boson}},  {\em Nucl. Phys.} {\bf B671} (2003) 148--174,
  [\href{http://xxx.lanl.gov/abs/hep-ph/0306259}{{\tt hep-ph/0306259}}].

\bibitem{Agashe:2004rs}
K.~Agashe, R.~Contino, and A.~Pomarol, {\it {The Minimal Composite Higgs
  Model}},  {\em Nucl. Phys.} {\bf B719} (2005) 165--187,
  [\href{http://xxx.lanl.gov/abs/hep-ph/0412089}{{\tt hep-ph/0412089}}].

\bibitem{Giudice:2007fh}
G.~F. Giudice, C.~Grojean, A.~Pomarol, and R.~Rattazzi, {\it {The
  Strongly-Interacting Light Higgs}},  {\em JHEP} {\bf 06} (2007) 045,
  [\href{http://xxx.lanl.gov/abs/hep-ph/0703164}{{\tt hep-ph/0703164}}].

\bibitem{Bryan:1993hz}
J.~A. Bryan, S.~M. Carroll, and T.~Pyne, {\it {A Texture bestiary}},  {\em
  Phys. Rev.} {\bf D50} (1994) 2806--2818,
  [\href{http://xxx.lanl.gov/abs/hep-ph/9312254}{{\tt hep-ph/9312254}}].

\bibitem{Hill:2007zv}
C.~T. Hill and R.~J. Hill, {\it {$T^-$ parity violation by anomalies}},  {\em
  Phys. Rev.} {\bf D76} (2007) 115014,
  [\href{http://xxx.lanl.gov/abs/0705.0697}{{\tt arXiv:0705.0697}}].

\bibitem{Murayama:2009nj}
H.~Murayama and J.~Shu, {\it {Topological Dark Matter}},  {\em Phys. Lett.}
  {\bf B686} (2010) 162--165, [\href{http://xxx.lanl.gov/abs/0905.1720}{{\tt
  arXiv:0905.1720}}].

\bibitem{Joseph:2009bq}
A.~Joseph and S.~G. Rajeev, {\it {Topological Dark Matter in the Little Higgs
  Models}},  {\em Phys. Rev.} {\bf D80} (2009) 074009,
  [\href{http://xxx.lanl.gov/abs/0905.2772}{{\tt arXiv:0905.2772}}].

\bibitem{Gillioz:2010mr}
M.~Gillioz, A.~von Manteuffel, P.~Schwaller, and D.~Wyler, {\it {The Little
  Skyrmion: New Dark Matter for Little Higgs Models}},  {\em JHEP} {\bf 03}
  (2011) 048, [\href{http://xxx.lanl.gov/abs/1012.5288}{{\tt
  arXiv:1012.5288}}].

\bibitem{Trodden:2004ea}
M.~Trodden and T.~Vachaspati, {\it {Topology in the little Higgs models}},
  {\em Phys. Rev.} {\bf D70} (2004) 065008,
  [\href{http://xxx.lanl.gov/abs/hep-ph/0404105}{{\tt hep-ph/0404105}}].

\bibitem{Auzzi:2006ns}
R.~Auzzi and M.~Shifman, {\it {Low-Energy Limit of Yang-Mills with Massless
  Adjoint Quarks: Chiral Lagrangian and Skyrmions}},  {\em J. Phys.} {\bf A40}
  (2007) 6221--6238, [\href{http://xxx.lanl.gov/abs/hep-th/0612211}{{\tt
  hep-th/0612211}}].

\bibitem{Bolognesi:2007ut}
S.~Bolognesi and M.~Shifman, {\it {The Hopf Skyrmion in QCD with Adjoint
  Quarks}},  {\em Phys. Rev.} {\bf D75} (2007) 065020,
  [\href{http://xxx.lanl.gov/abs/hep-th/0701065}{{\tt hep-th/0701065}}].

\bibitem{Auzzi:2008hu}
R.~Auzzi, S.~Bolognesi, and M.~Shifman, {\it {Skyrmions in Yang--Mills Theories
  with Massless Adjoint Quarks}},  {\em Phys. Rev.} {\bf D77} (2008) 125029,
  [\href{http://xxx.lanl.gov/abs/0804.0229}{{\tt arXiv:0804.0229}}].

\bibitem{Bolognesi:2009vm}
S.~Bolognesi, {\it {Skyrmions in Orientifold and Adjoint QCD}},
  \href{http://xxx.lanl.gov/abs/0901.3796}{{\tt arXiv:0901.3796}}.

\bibitem{Balachandran:1982ty}
A.~P. Balachandran, V.~P. Nair, N.~Panchapakesan, and S.~G. Rajeev, {\it {Low
  Mass Solitons from Fractional Charges in QCD}},  {\em Phys. Rev.} {\bf D28}
  (1983) 2830.

\bibitem{Balachandran:1983dj}
A.~P. Balachandran, A.~Barducci, F.~Lizzi, V.~G.~J. Rodgers, and A.~Stern, {\it
  {A Doubly Strange Dibaryon in the Chiral Model}},  {\em Phys. Rev. Lett.}
  {\bf 52} (1984) 887.

\bibitem{Derrick:1964ww}
G.~H. Derrick, {\it {Comments on nonlinear wave equations as models for
  elementary particles}},  {\em J. Math. Phys.} {\bf 5} (1964) 1252--1254.

\bibitem{Braaten:1988cc}
E.~Braaten and L.~Carson, {\it {The deuteron as a toroidal skyrmion}},  {\em
  Phys. Rev.} {\bf D38} (1988) 3525.

\bibitem{Adkins:1983nw}
G.~S. Adkins and C.~R. Nappi, {\it {Stabilization of Chiral Solitons via Vector
  Mesons}},  {\em Phys. Lett.} {\bf B137} (1984) 251.

\bibitem{Jackson:1985yz}
A.~Jackson, A.~D. Jackson, A.~S. Goldhaber, G.~E. Brown, and L.~C. Castillejo,
  {\it {A modified skyrmion}},  {\em Phys. Lett.} {\bf B154} (1985) 101--106.

\bibitem{Bogomolny:1975de}
E.~B. Bogomolny, {\it {Stability of Classical Solutions}},  {\em Sov. J. Nucl.
  Phys.} {\bf 24} (1976) 449.

\bibitem{Bott:1956tf}
R.~Bott, {\it {An Application of Morse theory to the topology of Lie groups}},
  {\em Bull. Soc. Math. Fr.} {\bf 84} (1956) 251--281.

\bibitem{Faddeev:1996zj}
L.~D. Faddeev and A.~J. Niemi, {\it {Knots and particles}},  {\em Nature} {\bf
  387} (1997) 58, [\href{http://xxx.lanl.gov/abs/hep-th/9610193}{{\tt
  hep-th/9610193}}].

\end{thebibliography}\endgroup
\bibliographystyle{JHEP}

\end{document}